\documentclass[sn-mathphys]{sn-jnl}

\jyear{2022}%
\usepackage{pbox}

\theoremstyle{thmstyleone}%
\theoremstyle{thmstyletwo}%

\theoremstyle{thmstylethree}%

\begin{document}

\title[Local Change Point Detection and  Cleaning of EEMD Signals]{Local Change Point Detection and  Cleaning of EEMD Signals}
\author*[1]{\fnm{Kentaro} \sur{Hoffman}}\email{kjh4@live.unc.edu}

\author[2]{\fnm{Jonathan} \sur{Lees}}

\author[1]{\fnm{Kai} \sur{Zhang}}

\affil*[1]{\orgdiv{Department of Statistics and Operations Research}, \orgname{University of North Carolina at Chapel Hill}, \orgaddress{\street{150 E. Cameron Ave}, \city{Chapel Hill}, \postcode{27514}, \state{NC}, \country{USA}}}

\affil*[2]{\orgdiv{Department of Earth, Marine, Environmental Sciences}, \orgname{University of North Carolina at Chapel Hill}, \orgaddress{\street{104 South Road}, \city{Chapel Hill}, \postcode{27514}, \state{NC}, \country{USA}}}

\abstract{The Ensemble Empirical Mode Decomposition (EEMD) has become a preferred technique to decompose nonlinear and non-stationary signals due to its ability to create time-varying basis functions. However, current EEMD signal cleaning techniques are unable to deal with situations where a signal only occurs for a portion of the entire recording length. By combining change point detection and statistical hypothesis testing, we demonstrate how to clean a signal to emphasize unique local changes within each basis function. This not only allows us to observe which frequency bands are undergoing a change, but also leads to improved recovery of the underlying information. Using this technique, we demonstrate improved signal cleaning performance for acoustic shockwave signal detection. The technique is implemented in R via the \texttt{\texttt{LCDSC}}  package.}

\keywords{Change Point Detection \sep Signal Cleaning \sep EEMD \sep Sparsity  \sep \texttt{\texttt{LCDSC}}}
\maketitle

\section{Introduction}
\label{sec:introduction}
The Ensemble Empirical Mode Decomposition (EEMD) method has become an important technique for the decomposition of nonlinear and non-stationary signals in fields including medicine \cite{sleep1,medicine1}, hydrology \cite{water}, seismology \cite{siesmology}, and mechanical engineering \cite{gyro,eng1}. A reason for its success has been EEMD's ability to create data-adaptive, rather than predefined, basis functions called Intermediate Mode Functions (IMFs). These adaptive basis functions can be non-stationary and nonlinear, making them ideal for complex signals that are not as natural to express in Fourier or Wavelet bases. 

\begin{figure}
    \centering
    \includegraphics[scale = 0.35]{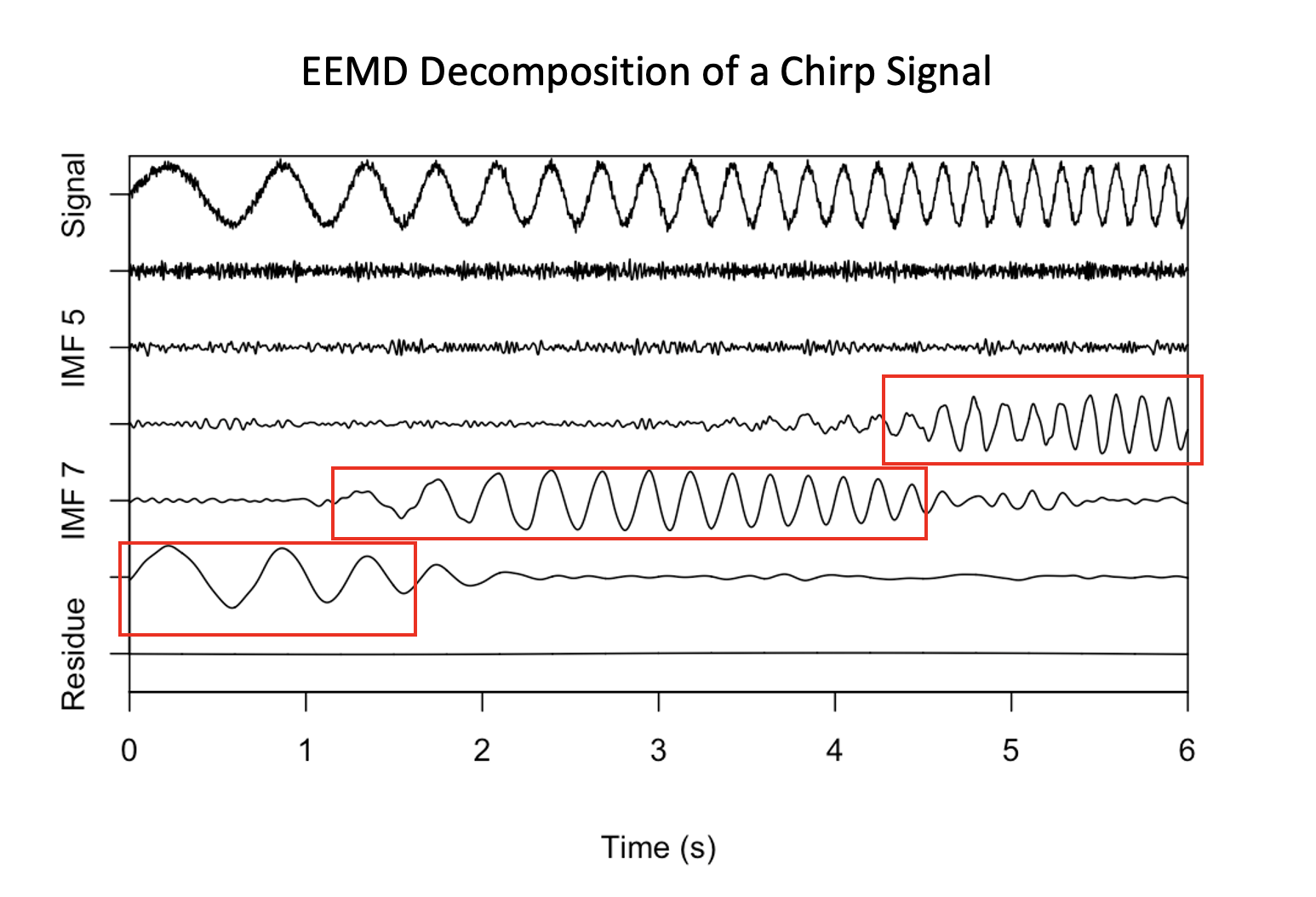}
    \label{fig:chirp3}
    \caption{ A chirp with white noise decomposed by EEMD. The boxed-in areas identify when each basis function is picking up the sinusoidal signal. Notice how the increasing frequency of the sinusoid makes it such that no basis function picks up the signal for the entire duration. }
\end{figure}

However, this data-adaptive nature of the EEMD's basis functions can make it hard to know \textit{a priori} in which basis function a signal may end up. For instance, consider a chirp signal linearly increasing in frequency perturbed with white noise. When decomposed by EEMD, we can see in Figure \ref{fig:chirp3} that the signal glides between IMFs 8-6. 
Common EEMD signal cleaning techniques such as those used in \cite{high1, high2, band1, band2,cor1, cor2,length1, white1, white2} first decompose the signal into its base IMF functions, but then treat the entire length of an IMF as either signal or noise. However, in this example, due to the increasing frequency of the chirp signal, no basis function is consistently signal or noise. To properly clean this signal, a more nuanced technique that is able to identify subsections of IMF as signal or noise is necessary. In this paper, we provide a novel example of an EEMD signal cleaning technique, Local Change Detection and Signal Cleaning (\texttt{\texttt{LCDSC}}), that is able to identify and clean subsections of EEMD signals. Moreover, we show how this technique can improve the identification of acoustic shock waves.

\section{Local Change Point Detection and Signal Cleaning }
\subsection{EEMD}

The Empirical Mode Decomposition (EMD) was invented in 1998 as a novel technique for analyzing nonlinear and non-stationary time series data \cite{Huang1998}. One of the most popular varients of this method is the Ensemble Empirical Mode Decomposition (EEMD)  \cite{Wu2009EnsembleEM} . The EEMD was invented  \textit{et al.} uses iteratively computed, adaptive filters to decompose a signal $X(t)$ into
\begin{equation}
X(t)  = \sum_{j=1}^n IMF_j(t) + r(t).
\end{equation}
Here, $IMF_j(t)$ is the $j$-th basis function, referred to as an Intermediate Mode Function (IMF), and $r(t)$ is the residual. As the EEMD is a numerical algorithm, there exists a variety of stopping criteria to indicate when the algorithm has converged. While many stopping criteria exist for the EEMD, one of the most common, called S-stoppage, \cite{Swave} results in the remainder term $r(t)$ becoming a monotonic or a constant function. In either case, the resulting $r(t)$ can easily be subtracted from the original signal $X(t)$ to create a decomposition with no residual term. Thus for the purposes of this paper, we will assume that either $X(t)$ has an $r(t)$ of 0, or $X(t)$ has had its remainder subtracted out resulting in
\begin{equation}
X(t)  = \sum_{j=1}^n IMF_j(t).
\end{equation}
Using the Hilbert Transform, each $IMF_j$'s instantaneous amplitude $a_j(t)$ and instantaneous frequency  $w_j(t)$  time series can be extracted as the sum of a time-varying amplitude function $a_j(t)$ multiplied by an equally time-varying frequency function $e^{i w_j(t)}$,
\begin{equation}
X(t)  = \sum_{j=1}^n IMF_j(t)  \\ 
=\sum_{j=1}^n a_j(t) e^{i  w_j(t)} .
\label{eq:analytic}
\end{equation}
As $a_j(t)$ and $w_j(t)$ are functions of time, this decomposition allows for the analysis of time varying amplitude and frequency signals. This contrasts with the Fourier decomposition in which the amplitude $a_j$ and instantaneous frequency $w_j$ are no longer functions of time, but constants. 

IMFs also come with several crucial properties. By definition, an IMF is a nonlinear oscillatory function that satisfies the requirements \cite{hht}:
\begin{enumerate}
    \item For each IMF, the number of local extrema and zero crossings must differ by at most one.
    \item Let $g_{j,max}(t)$ and $g_{j,min}(t)$ be smooth functions connecting the local maxima and minima of the $j$-th IMF (These functions are commonly referred to as the upper and lower envelope of $X(t)$). At any time point $t$, the mean of the upper envelope of the $j$-th IMF, $g_{j,max}(t)$, and the lower envelope, $g_{j,min}(t)$, is zero:
    \begin{equation}
        g_{j,max}(t) + g_{j, min}(t) = 0.
    \end{equation}
\end{enumerate}
Moreover, \cite{hhtbook} showed that IMFs are approximately orthogonal to one another, allowing for the decomposition:

\begin{equation}
     X(t)^2   = \sum_{j=1}^n IMF_j(t)^2  \\ 
   = \sum_{j=1}^n a_j^2(t) e^{2iw_j(t)}.
\end{equation}

Finally, \cite{hhtbook} illustrated that under white noise, the power distribution within each IMF is approximately normally distributed. These properties are crucial in crafting a powerful change point detection and signal cleaning algorithm.

\subsection{Additive Local Noise Model}
In performing local change point detection, we will operate under the assumption that there exists an observed signal $X(t)$ that consists of Gaussian noise $R(t)$ occurring throughout the entire duration and an underlying true signal $S(t)$ which is only observable during the interval $A$. If we assume an additive decomposition this gives the setup
\begin{equation}
    X(t) = S(t) I_{A}(t) + R(t),
\end{equation}
where $I_{A}(t)$ is an indicator function that returns 1 if $t \in A$ and 0 otherwise. The additional assumption of statistical independence between $R(t)$ and both $S(t)$ and the set $A$ completes the additive local noise model.
\subsection{Change Point Detection of the IMFs}Under the additive local noise model, the goal of signal cleaning is to recover the true signal $S(t)$ by first estimating the interval $A$, or when the true signal is occurring, and then performing a signal cleaning on $X(t)$ for $t \in A$ to recover $S(t)$. To identify when changes are occurring in $X(t)$, we first decompose $X(t)$ into its constituent IMFs and then perform a change point detection procedure on each IMF. Here, IMF is the ensembeled IMF from the runs of the EEMD. From a statistical perspective, identifying change points entails finding the set of time points $\{\tau_1^{(i)}, ...,\tau_{n_j}^{(i)} \}$ such that:
\begin{equation}
\begin{aligned}
    f(IMF_j (t_1)) \neq f(IMF_j(t_2)), \\
    \forall t_1 \in [\tau_k^{(i)}, \tau_{k+1}^{(i)}], \\
    \forall t_2 \in (\tau_{k+1}^{(i)}, \tau_{k+2}^{(i)}],\\
    \forall k \in [1, ..., k-2].
\end{aligned}
\end{equation}
Here, $f(IMF_j(t))$ represents the distribution of the ensembled $IMF_j$ at time $t$. However, as the distribution of each IMF is generally unknowable \textit{a priori} outside of well-known distributions such as white noise \cite{WANG2013581}, it can be difficult to create an change point detection algorithm that is able to rapidly identify when a change is occurring. To make this more tractable, we utilize several of the properties of IMFs and the additive local noise model to construct a more feasible change point detection problem.

According to our additive local noise model
\begin{equation}
    X(t)^2 = 
     \begin{cases}
       (S(t)  + R(t))^2 & \text{ If } t \in A \\
       R(t) ^2 & \text{ If } t \not \in A. \\
     \end{cases}
\end{equation}
Combining this with the statistical independence between $R(t)$, $S(t)$ and $A$ we assumed in the additive local noise model, this implies that
\begin{equation}
    E[X(t)^2] = 
     \begin{cases}
       E[S(t)^2]  + E[R(t)^2] & \text{ If } t \in A \\
       E[R(t) ^2] & \text{ If } t \not \in A. \\
     \end{cases} 
\end{equation}
Thus, when we are in interval $A$, there is an increase in expected power in $X(t)$ (power being $X(t)^2$). Furthermore, by the orthogonality of the IMFs, this directly implies that an increase in power in $X(t)$ must lead to a corresponding increase in at least one of the constituent IMFs. Formally, if $t \in A$, then there exist a subset of IMFs $\eta \subset \{1, ... , n\}$ such that for $j \in \eta$, $IMF_j(t)$ displays an increase in power during $t \in A$. 
\begin{equation}
    \forall t \in A, t^* \not \in A, \exists j \in \eta \neq \emptyset : E(IMF_j(t)^2 ) \geq E(IMF_j(t^*)^2)
\end{equation}
Additionally, as each IMF has a mean of zero with respect to its envelope, an increase in power in an IMF implies an increase in the variance in that IMF
\begin{equation}
    E[IMF_j(t)^2] = E[(IMF_j(t) - E(IMF_j(t))^2] = Var(IMF_j(t)).
\end{equation}
Using equation \ref{eq:analytic}, we can write the variance of an IMF as:
\begin{equation}
    Var(IMF_j(t)) = Var(a_j(t) e^{i w_j(t)})
\end{equation}
where $|a_j(t)|$ is the instantaneous amplitude and $\frac{1}{2 \pi } \frac{d w_j(t)}{dt}$ is the instantaneous frequency. If the variance of an IMF increases, this could be because $a_j(t)$ has changed or because $e^{i w_j(t)}$ has changed. However, in simulation, we observe that changes in the variance of an IMF are better expressed in the amplitude term, $a_j(t)$ rather than the frequency term, $e^{i w_j(t)}$. Thus, to identify a local signal, we will look for IMFs which are exhibiting a change in the variance of their amplitudes. 

\subsection{Change Point Detection}
To identify when the amplitude of an IMF is experiencing an increase in variance, we employ techniques from a well-developed branch of statistics, change point detection. Many change point detection problems can be framed in the form of minimizing an objective function of the form:
\begin{equation}
  \min_{m} \min_{\tau_1, ... \tau_{m-1}} \sum_{i=1}^{m-1} L(X_{\tau_{i-1}  },  X_{ \tau_{i}}, X_{ \tau_{i+1} -1} ) + \beta  D(m),  
\end{equation}
where $\tau_0$ is 1 and $\tau_m$ is the length of the signal, $m$ is the number of change points, $\tau_i$ is the location of the $i$-th change point, $\beta$ is a constant, $L$ is a function that decreases when $\tau$ is a true change point, and $D(m)$ is a penalization function that increases with the number of change points selected. By balancing $L$ and $D(m)$, the objective seeks to select the correct number and locations of changes in variance. 

For our particular type of local signal, since the background noise is Gaussian, we require an L and D(m) that is well suited to noticing changes in Gaussian signals. One such L is the likelihood ratio test for changes in variance of Gaussians \cite{cp}.
\begin{equation}
    L(X_{\tau_{i-1}  },  X_{ \tau_{i}}, X_{ \tau_{i+1}-1} )  =   \frac{C_{\tau_{i} } } {C_{\tau_{i+1} -1}} - \frac{\tau_{i} - \tau_{i-1}}{\tau_{i+1} -1 - \tau_{i-1}}
\end{equation}
Here $C_{\tau_i}$ is the cumulative normalized second moment, $ \sum_{k= {\tau_{i-1}+1}}^{{\tau_{i}}} (X(k) - \overline{X_{ \tau_i}})^2$ and $\overline{X_{ \tau_i}}$ is the cumulative mean, $ 1/(\tau_i - \tau_{i-1} +1) \sum_{k=\tau_{i-1} +1}^{\tau_i } X(k) $. This L has the ability not only to consistently select the correct location, but correct number of change points under an asymptotic scheme but also has strong performance in the finite sample case \cite{cp}. As for $\beta D(m)$, this is a penalization term that combines some function of the number of change points, $D(m)$, with a constant, $\beta$ to ensure that the correct number of change points are selected \cite{TRUONG2020107299}. Whiile there exist many popular penalty terms such as Akaike's Information Criterion ($  \beta m$)  \cite{aic} and Bayesian Information Criterion  ($m \log(n)$) \cite{bic} ($n$ is the total signal length), many still lack theoretical justifications in the context of change point detection. One exception is the  newer Modified Bayesian Information Criterion $(-1/2  (3 m + \log(n) + \sum_{i=1}^{m+1} \log(\tau_{i} - \tau_{i-1} )) $ which by a creative large sample approximation of a Bayes Factor, provides a solid theoretical justification to select the correct number and location of Gaussian change points \cite{mbic}.

\subsection{Hypothesis Test and Sparse Basis Selection}
Once the change points algorithm has identified the points where each IMF has undergone a change, we must determine if each interval between change points is either signal or noise. We propose a simple hypothesis-test-based algorithm that is able to automatically create sparsely cleaned IMFs based on changes in power. To identify which of the intervals have a statistically larger variance than their neighboring intervals, we employ the hypothesis test:
\begin{align}
    H_0: \sigma_{during}^2 \leq  \gamma * \max(\sigma_{before}^2, \sigma_{after}^2) \nonumber  \\
    H_1: \sigma_{during}^2  >  \gamma*\max(\sigma_{before}^2, \sigma_{after}^2 ),
    \label{eq:hypo}
\end{align}

where $\sigma_{before}^2$ is the variance of the previous interval, $\sigma_{during}^2$ is the variance of the current interval, $\sigma_{after}^2$ is variance of following interval, and $\gamma$ is assumed to be greater than or equal to 1.

By rearranging the alternate hypothesis, $ \gamma >   \frac{\sigma_{during}^2}{\max(\sigma_{before}^2, \sigma_{after}^2 )} $, we can see that $\gamma$ serves as a measure of how much the ratio of variances much increase to be considered significant. Setting $\gamma = 1$ tests if there has been any statistically significant increase in variance. A common test statistic for \eqref{eq:hypo} is the F-statistic for change in variance is:
\begin{equation}
    F_{before/during} = \frac{ \gamma*\max(S_{before}^2, S_{after}^2) }{S_{during}^2},
\end{equation} 
where $S_{before}^2$ is the sample variance used to estimate $\sigma_{before}$. $F_{before/during}$ is compared against the F distribution with degrees of freedoms, $df_1 = n_{during}$, $df_2 = max(n_{before}, n_{after})$ (where $n_{during}$ is the length of the during interval) to determine the p-value and thus significance. 

As this process involves performing a hypothesis test at every potential change point, across every IMF, this can quickly lead to a large number of tests being performed for the same goal: identifying a significant segment. This large number of tests can lead to the multiplicity issue where one or more spurious false positives may occur. To perform these  tests so they collectively have an $\alpha$ ($1 > \alpha > 0$) probability of a false positive (which is known as the the Family-Wise Error Rate), we employ the multiple testing correction method, Holm-Bonferroni method \cite{bonf}.
If the p-value for a segment is significant after the Holm-Bonferroni correction, then we can claim that the interval contains the desired signal. If not, the interval does not contain the true signal. If an IMF only has one change point (and thus cannot have a before, during, and after interval), then then $\max(\sigma_{before}^2, \sigma_{after}^2)$ is replaced with $\sigma_{after}^2$.
\subsubsection{Signal Cleaning}
Once the significant segments are identified, we must determine how a signal is cleaned. Similar to how many Fourier-domain filters clean signals by setting some Fourier bases to zero, in EEMD signal cleaning, there exist similar methods based on basis removal (See Table 1 for examples). In this spirit, our signal cleaning algorithm will set a signal segment to 0 if it is not identified as containing a significant signal according to the previous hypothesis test. If there are no change points in an IMF, then there is no identifiable local signal, only noise, so the entire IMF is set to 0. We note here that if a segment of an IMF is considered significant, that segment is included in its entirety. It is likely that one can further improve the performance by including a smoothing or SURE based cleaning procedure to clean the significant segments in addition to setting nonsignificant segments to zero. However, to do this, careful work must be put into determining the appropriate level of cleaning for each IMF, a nontrivial question. 

\begin{figure}
    \centering
    \includegraphics[width = \textwidth]{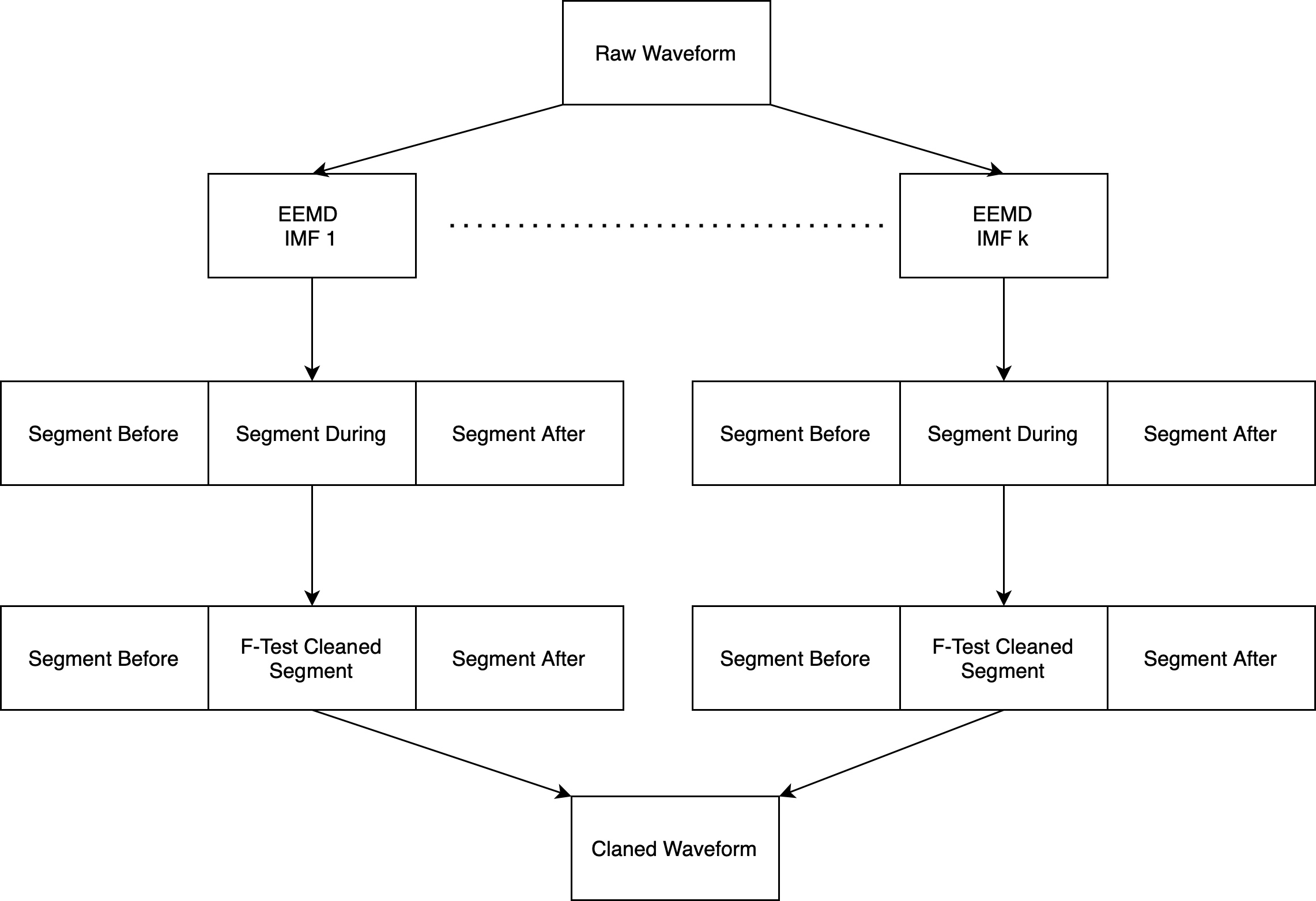}
    \caption{Flowchart of the LCDSC algorithm}
    \label{fig:my_label}
\end{figure}

\section{Simulation}
\subsection{Simulation 1: Doppler Signal}
\begin{figure}[h]
    \centering
    \includegraphics[scale = 0.35]{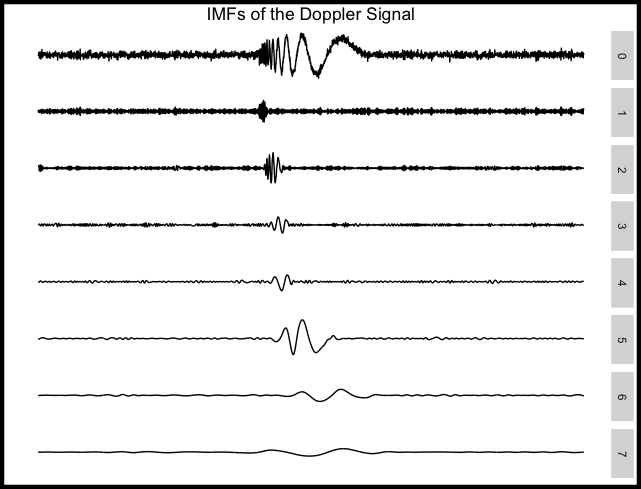}
    \caption{EEMD of the Local Doppler Signal. The IMF numbers are on the right with ``IMF 0" referring to the original signal. In the EEMD, none of the IMFs are purely signal or noise necessitating a local signal cleaning procedure. }
    \label{fig:b1}
\end{figure}

\begin{figure}[h]
    \centering
    \includegraphics[scale = 0.35]{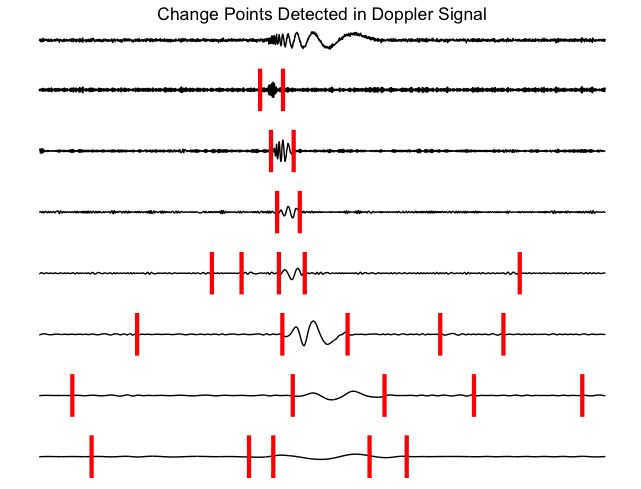}
    \caption{Change points that were detected in the Local Doppler Signal in Figure \ref{fig:b1} when employing the normal likelihood ratio objective function and the Modified Bayesian Information Criterion over-fitting penalty. }
    \label{fig:b2}
\end{figure}

\begin{figure}[h]
    \centering
    \includegraphics[scale = 0.35]{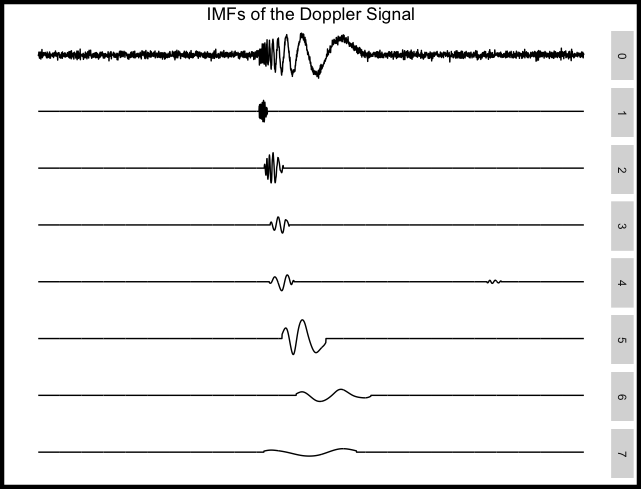}
    \caption{The IMFs in Figure \ref{fig:b2} after each section that was identified by the change point detection algorithm was cleaned using the F-test/Hole-Bonferroni procedure with $\gamma = 1$. Notice how the basis functions are set to 1 when the signal is not present within the basis function.  }
    \label{fig:b3}
\end{figure}

\begin{figure}[!htbp]
    \centering
        \includegraphics[scale = 0.34]{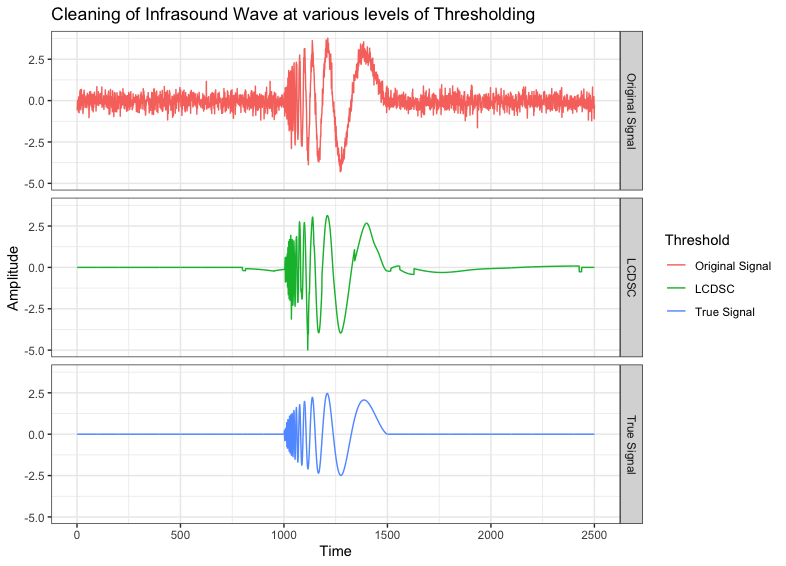}
    \caption{Comparison of the original signal with the cleaned signal. The \texttt{LCDSC} recovers much of the original signal. It  performs especially well at cleaning the signal to closely match the true start and end points.  }
    \label{fig:b4}
\end{figure}

To demonstrate this signal cleaning procedure, we take a synthetic example where a Doppler signal is hidden in the midst of Gaussian white noise. The Doppler is a classic example of a nonlinear signal with variable frequency, exactly the kinds of signals that the flexible EMD algorithm is well suited for. We will refer to this as the Local Doppler example.  For Simulation 1, we will use a Local Doppler of length 2500 with the Doppler occurring during the middle of the signal:
\begin{equation}
    X(t) = \begin{cases}
		R(t)  & \mbox{if } t < 1000  , t > 1500 \\
	    S(\frac{t - 1000}{1500}) + R(t) & \mbox{if } 1000 \leq t \leq 1500.
	\end{cases}
\end{equation}
$S(t)$ is the Doppler signal from \cite{10.1093/biomet/81.3.425} rescaled to occur between [1000,1500]: 
\begin{equation}
S(t) = 7(t (1-t)^{0.5} \sin (2 \pi (1+ 0.05)/(t+ 0.05)).    
\end{equation}
The goal of the this simulation would be to have the algorithm:
\begin{itemize}
    \item Identify when the signal started and ended (time points :1000-1500)
    \item Clean the Signal that was isolated
\end{itemize}
As can be seen in Figure \ref{fig:b1}, the Doppler signal in the middle is expressed in all 7 IMFs with first IMFs expressing the higher frequency parts of the signal and the latter IMFs expressing the lower frequency sections. Moreover, no single IMF is ever purely signal or purely noise necessitating a local change point detection and signal cleaning. 
Running the change point detection algorithm in Figure \ref{fig:b2} at an $\alpha = 0.05$ type I error level and $\gamma = 1$ identifies many locations at which a change in the signal was detected. While IMFs 1-3 correctly identify two changes, one when the Doppler signal starts within their IMF and one when it ends, in IMFs 4-7, many spurious change points are detected that are not necessarily due to the Doppler signal. To remove these, the F-test cleaning step is performed. 
The resulting cleaned signal in Figures \ref{fig:b3} and \ref{fig:b4} illustrates how all of the change points outside of the duration of the Doppler signal were deemed nonsignificant by Holm-Bonferroni and set to zero. Not only does this provide a good estimation of the shape of the Doppler signal, matching the general sinusoidal shape and increasing frequency, but \texttt{LCDSC} provides a good estimate of when the Doppler signal starts, as the first nonzero point in IMF1 is at point 1010, only 1\% of the way into the start of the Doppler signal.

\subsection{Simulation 2: Doppler Signal-Comparison Study}
To compare the performance of our algorithm, we  extend our Doppler simulation from Simulation 1 and compare our performance against other EEMD signal cleaning techniques. These techniques come in two general varieties. Techniques 2-5 in Table \ref{tab:comp} are based on identifying some subset of the IMFs as containing only noise and cleaning the signal by completely removing the noise IMFs. With some of the cleaning procedures, the user must pre-specify how many basis functions to set to zero or clean through a trial-and-error process. To account for any possible variability in performance due to these subjective judgements, we will come up with an upper-bound for the performance of each algorithm by computing the best possible set of IMFs for each of the algorithms in question. 

As for the Wavelet Hard Thresholding (WHT) and Wavelet Interval Thresholing (WIT) cleaning techniques, these are based on performing a Wavelet-like thresholding on each of the IMFs \cite{wavethresh}. These compute the base noise level within each IMF and perform a hard or soft thresholding if the IMF lies within the expected noise band. While this method does not suffer from a subjective choice of IMF removal, it assumes that the true signal occurs throughout the entire duration of the signal, leading to a biased estimation of the base noise level. 

The data model for the simulation will utilize the Local Doppler Model with the middle containing our desired signal but with the total signal length T at differing values:

\begin{align}
X(t) = \begin{cases}
		R(t)  & \mbox{if } t < \frac{2}{5}T  , t > \frac{3}{5}T \\
	    S(t) + R(t) & \mbox{if } \frac{2}{5}T \leq t \leq \frac{3}{5}T.
	\end{cases} 
	\end{align}
T is tested at 1000, 2000, and 2500 time steps. $R(t)$ will again be Gaussian white noise but with the noise level varying from 0.2 to 0.5. The cleaned signal is then compared to the underlying Doppler signal and error computed in terms of Residual Sum of Squares (RSS) as this corresponds to the total power difference between the estimated and the cleaned signal,
\begin{equation}
    RSS = \sum_{t=1}^T (X(t) - Cleaned(t))^2.
\end{equation}
At each level of noise and signal length, 20 replicates of the simulation were performed.

\begin{table}[h]
\centering
    \begin{tabular}{|c|c|}
    \hline
        \textbf{Cleaning Method} &\textbf{Description}  \\
        \hline
        \texttt{LCDSC} & Our Method\\ \hline
        
        k-Highest  & Removal all but the k-highest IMFs  \cite{high1} \\ \hline
        l-Lowest &  Removal all but the k-lowest IMFs \cite{high2}  \\ \hline
        k-Highest \&  l-Lowest  & Combination of k-Highest and l-Lowest \cite{band1, band2}  \\ \hline
        Power Set Cleaning & Perform a best subset selection over all possible subsets.  \\ \hline
        WHT &  Wavelet  Hard Thresholding each IMF \cite{wavethresh}  \\ \hline
        WIT &  Wavelet  Interval Thresholding each IMF \cite{wavethresh} \\ \hline
        No Cleaning  & No Signal cleaning \\ \hline
    \end{tabular}
    \caption{List of  EEMD Signal Cleaning Techniques}
    \label{tab:comp}
\end{table}

\begin{figure}[!htbp]
    \centering
    \includegraphics[scale = 0.34]{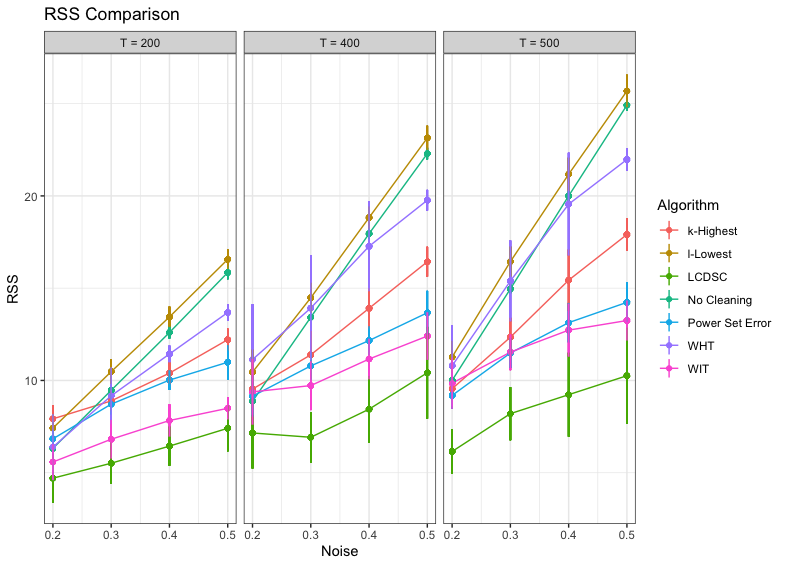}
    \caption{RSS Comparison of common cleaning methods vs \texttt{LCDSC}. The center point represents the mean RSS across 20 replicates and the bar represents one standard deviation from the center. From this, we observe that \texttt{LCDSC} performs better than competing signal cleaning methods, able to create the closest representation of the true signal. Note that this is RSS and not MSE so it is entirely expected that as the signal gets longer, the RSS should also increase.}
    \label{fig:comparison1}
\end{figure}

The results in Figure \ref{fig:comparison1} illustrate that across a wide scale of noise levels and sample sizes, the \texttt{LCDSC} performs well at local signal cleaning, uniformly outperforming other non-local signal cleaning techniques. 

\subsection{Simulation 3: Comparison Study- What if the signal is not local?}
While the \texttt{LCDSC} is built for the problem of local signal detection and cleaning, it is important to determine its performance as the duration of true signal is increased or decreased. We can express how local our signal is in terms of a ``locality Ratio":
\begin{align}
    \text{locality Ratio} = \frac{len(A)}{T - len(A)}.
\end{align}
$len(A)$ is the length of the interval A when the true signal is being expressed and T is the total length of the noisy signal. We vary the locality Ratio between 0 to 4, making the local signal cleaning problem increasingly local and favorable to \texttt{LCDSC}.

\begin{figure}[!htbp]
    \centering
    \includegraphics[scale = 0.1]{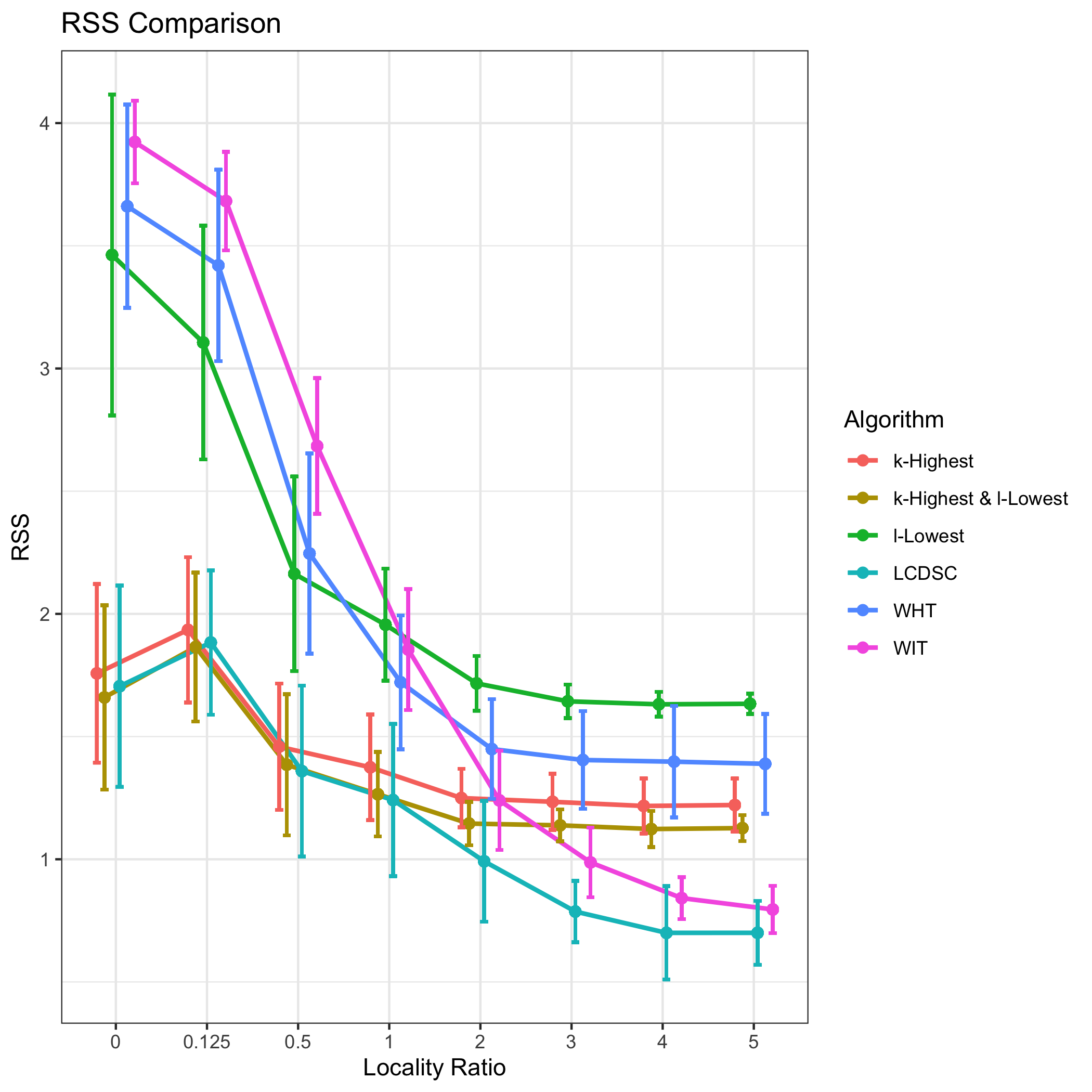}
    \caption{Changes in Residual Sum of Squares as the Locality Ratio is increased. When the noise ratio is low, the \texttt{LCDSC} performs slightly worse than k-Highest and k-Highest \& L-lowest, but once the noise ratio increases above 1, the \texttt{LCDSC} becomes the best performing method. No cleaning was not plotted as it had a much higher error than all the others and Power Set was near equivalent to k-Highest \& l-Lowest. }
    \label{fig:c2}
\end{figure}

Figure \ref{fig:c2} illustrates that when the locality Ratio is at or below one, then \texttt{LCDSC} performs approximately the same as the best performing method such as  k-Highest. However, once the locality ratio goes beyond one, \texttt{LCDSC} becomes the dominant signal cleaning technique followed by WIT. This gives us a rough guide for when to start considering a signal cleaning problem local or global. When the Noise Ratio is below one, it can be better to clean with global cleaning methods whereas local cleaning methods are better when the ratio is greater than one.

\subsection{Simulation 4: Additional Simulations - Distinguishing Consecutive Signals}
In previous simulations, we have focused our attention on examples where we have one true that is preceded and followed by white noise. However, because our algorithm makes no assumptions about the number of true signals, it is also useful in situations where we are interested in isolating multiple true signals. To demonstrate this, we will consider the situations where we have two Doppler signals separated by white noise of length $\delta$:

\begin{equation}
    X(t) =  \begin{cases}
  R(t,\sigma) & \text{ if } t < 500  \\
  S(\frac{t - 500}{1000}) + R(t,\sigma) & \text{ if } 500 \leq  t <  1000  \\
  R(t,\sigma) & \text{ if }  1000 \leq  t  < 1000 + \delta   \\
  S(\frac{t - 1000 + \delta}{1500 + \delta}) + R(t,\sigma) & \text{ if }  1000 + \delta  \leq  t  < 1500 + \delta   \\
  R(t,\sigma) & \text{ if }  1500 + \delta  \leq  t  <  2000 + \delta  
\end{cases}.
\end{equation}
Here $\delta \in \mathcal{N}$ controls the gap between the two Doppler signals and $R(t,\sigma) \sim N(0, \sigma^2)$ controls the standard deviation of the Gaussian background noise. As we will see as $\delta$ is decreased and $\sigma$ is increased, it will become progressively difficult to distinguish the Doppler signals from each other. 

\begin{figure}
    \centering
    \includegraphics[width = 0.8\textwidth]{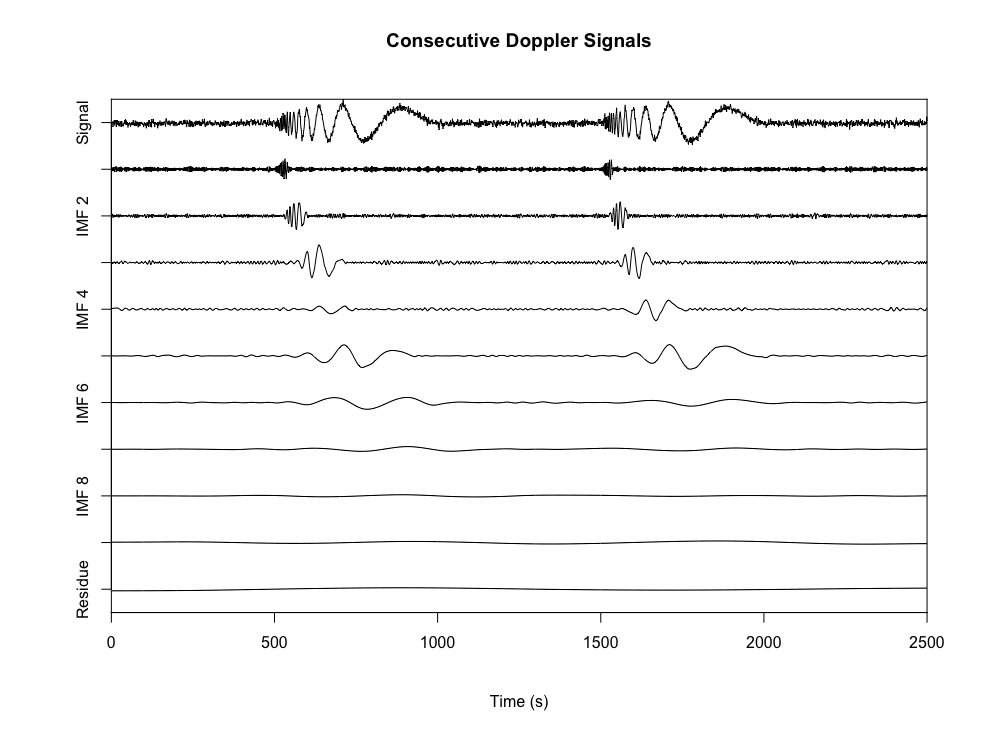}
    \includegraphics[width = 0.43\textwidth]{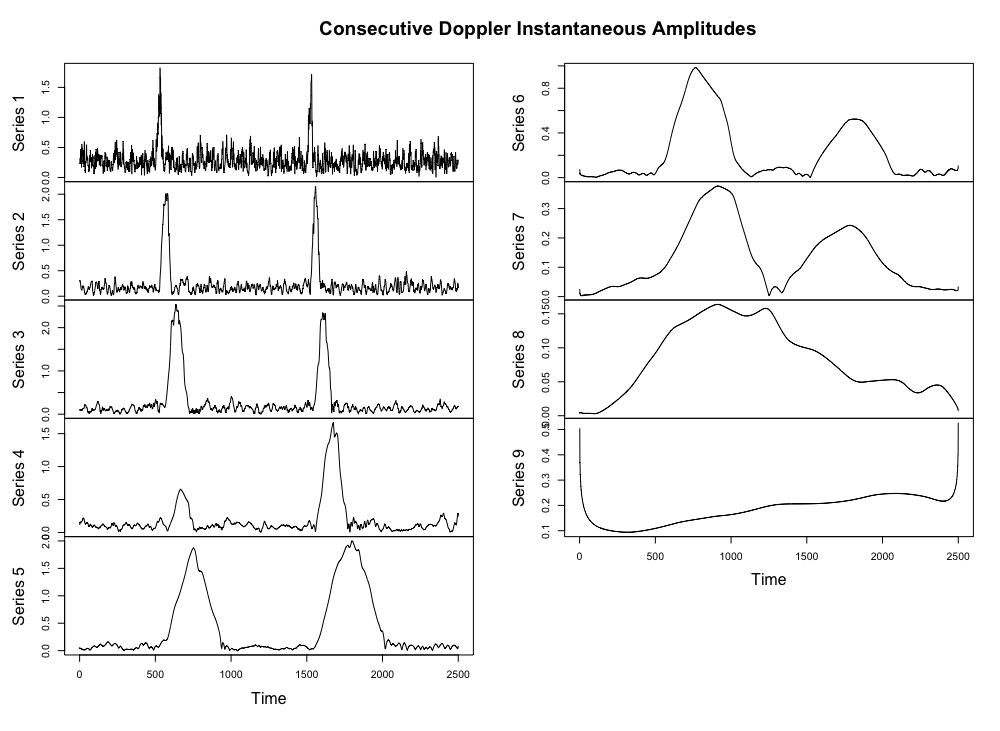}
    \includegraphics[width = 0.43\textwidth]{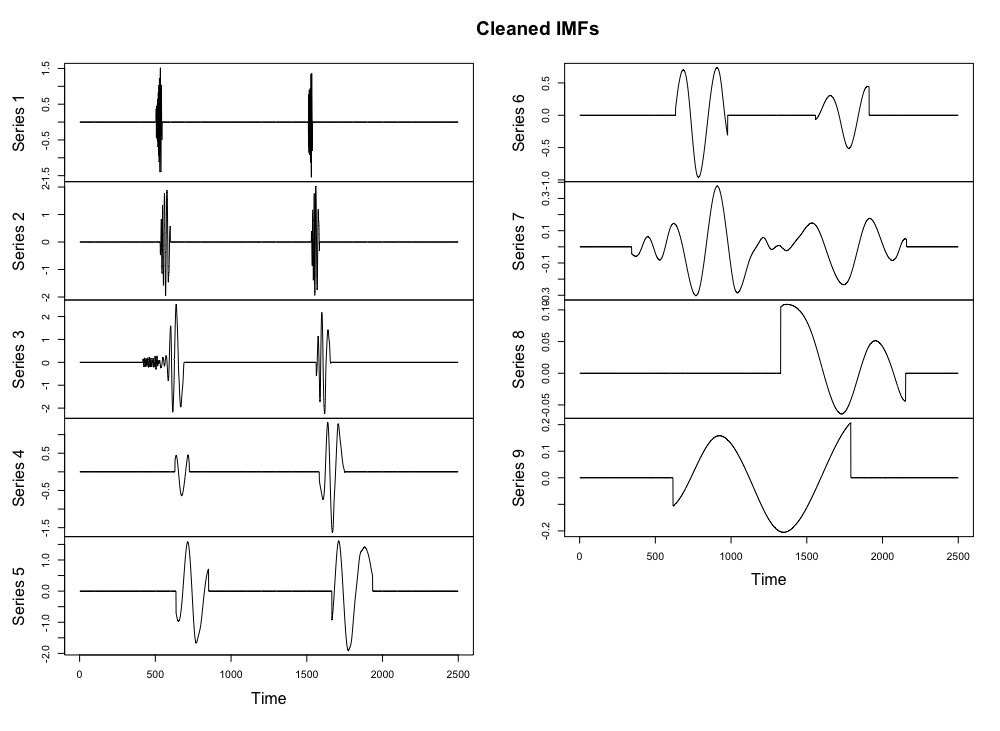}
    \caption{Top: Raw Signal and IMFs for the consecutive Dopplers (Bottom Left): Instantaneous Amplitudes for the IMFs. Note how at IMF9, we see stronger boundary effects in the instantaneous amplitude. In signals with more lower frequency components, this computation may require care. (Bottom Right): Cleaned IMFs   }
    \label{fig:two1}
\end{figure}

Looking at an example of such a signal when $(\delta = 500, \sigma = 0.25)$, we can see in Figure \ref{fig:two1} a plot of the IMFs, the instantaneous amplitudes, and the cleaned IMFs. From this, we can already notice several properties. 1) the increases in instantaneous amplitudes relative to the white noise is most apparent in the middle IMFs (3-5 for this example ). This is because the amplitude of the Doppler signal is highest in the middle frequencies. Thus we should expect the middle IMFs to be most distinguishable from background noise while the smallest and largest IMFs do no show clear spikes in amplitudes. 2) The Doppler signal is expressed at later and later time points as the IMF number increases. This is due to a direct property of the IMF decomposition and the Doppler Signal. Higher number IMFs express lower frequency signals and the Doppler signal increases in frequency over time. Thus, while they will not necessarily be occurring at the same time, there should still be two discernible spikes separated by a gap in the cleaned IMFs.

Thus for this simulation, we will be evaluating how many IMFs when cleaned yield two clear spikes with at least $50\%$ of the space in between, identified as noise and set to zero. So in the example of Figure \ref{fig:two1}, when $(\delta = 500, \sigma = 0.25)$, IMFs 1-6 exhibit the desired criteria while IMFs 7-9 do not. 

\begin{figure}
    \centering
    \includegraphics[width = \textwidth]{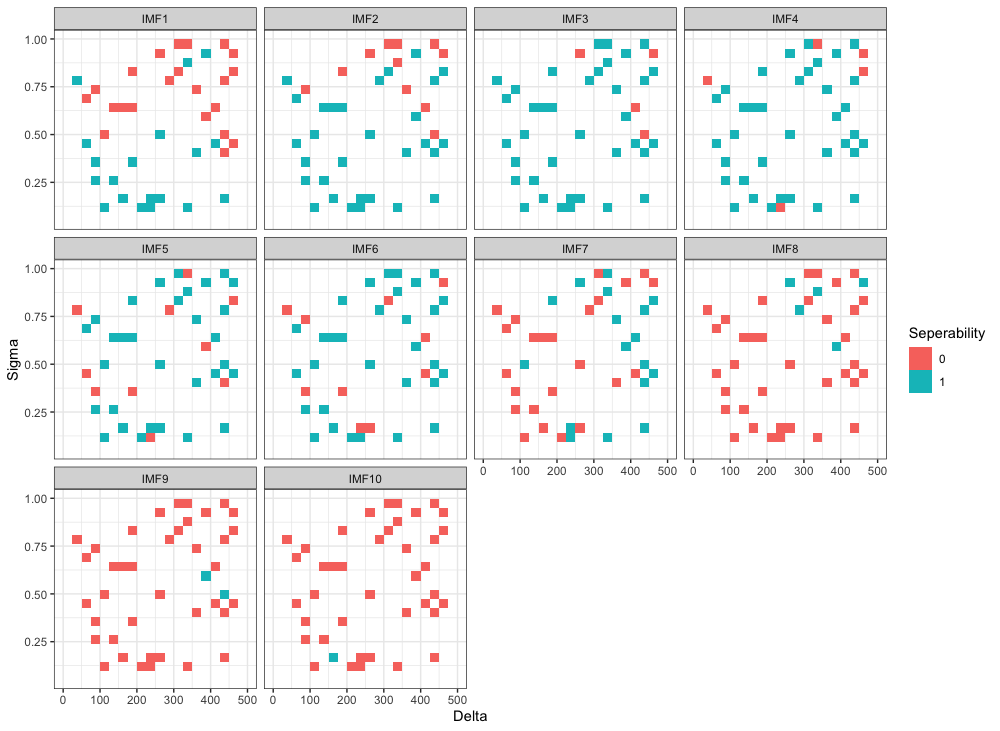}
    \label{fig:two2}
\end{figure}

In this simulation results in Figure \ref{fig:two2} , we have generated 50 signals with $\sigma$ drawn from Uniform[0.05,1] and $\delta$ from Uniform[10, 500]. These signals are then decomposed into their constituent IMFs and then each IMF is cleaned using our algorithm. If the IMF identifies, via human inspection, two clear spikes and more than $50\%$ of the space in between set to 0, we say that we have successfully cleaned both signals. The results of this separability study is shown in figure \ref{fig:two2}. Here we see that IMF1 is only separable when there is lower than 0.25 to 0.5 standard deviations of noise. Likewise IMF 2 also exhibits problems with separability when there is a high noise level, albiet with issues now occurring when above 0.75. IMFs 2-6 seem to be separable regardless of the level of background noise or gap size. But around IMF7-10, the separability of the IMFs seems to fall again, except unlike IMFs 1-2, the fall in separability seems to occur uniformly until only 1 or 2 out of 50 simulations show separable IMFs. From this, we can say that our algorithm into cleaning the signals into separable chunks depend on the level of background nose (espeically for high frequency IMFs), the IMF number, but is fairly robust to changes in the gap size. 

\section{Application}
\subsection{Application: Detection of Gliding events in Acoustic Explosions}

On October 28, 2014 an Anteras rocket operated by Orbital Sciences Corporation exploded shortly after takeoff \cite{nasa}. The resulting explosion was powerful enough that acoustic shockwave arrivals were observed at stations over 2000km away from the launch site. At the time, 226 acoustic and atmospheric stations from the Transportable USArray network were located within range of the explosion, resulting in arrivals from the explosion being picked up by the array's infrasound sensors. Many of these arrivals exhibited characteristics of dispersive waves at the infrasound level ($<$20 Hz). This is of interest as dispersive waves were only recognized recently in the infrasound domain \cite{waveguides} and because the Anteras explosion was one of the largest demonstrations to date of the existence of infrasound dispersive waves \cite{vergoz}. These dispersive waves are a result of the arrivals being reflected at different heights in the troposphere as well as being influenced by atmospheric conditions such as temperature and windspeed. This makes studying infrasound arrivals important tools in evaluating atmospheric density models \cite{vergoz}.

Isolating these dispersive waves can be complicated due to the relatively short time periods when the explosion was detected as well as the complex weather and atmospheric factors affecting recording conditions at each sensor.

\begin{figure}[ht]
    \centering
    \includegraphics[scale = 0.3]{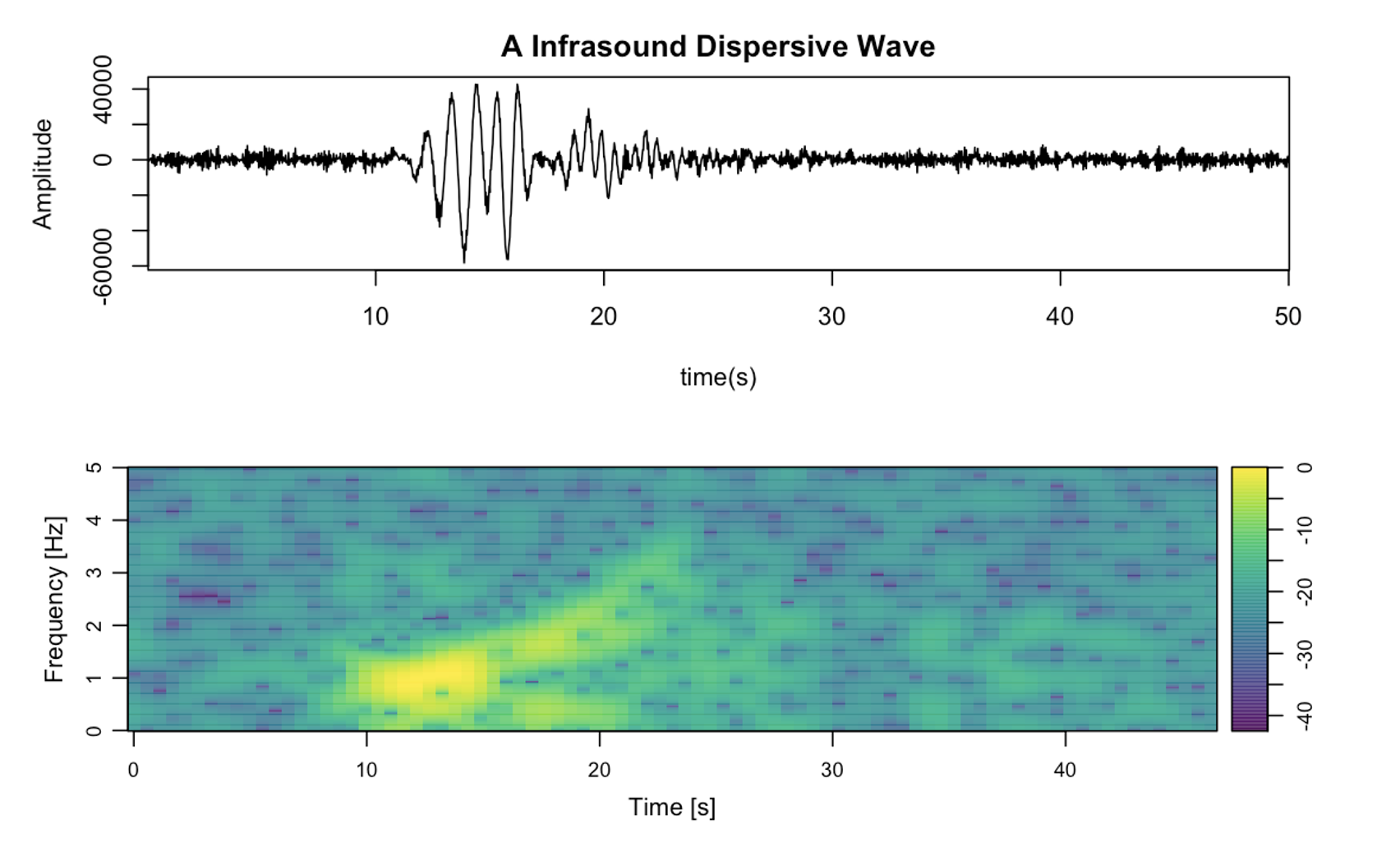}
    \caption{A canonical example of an Infrasound Dispersive wave. Note the increase in frequency over the duration of the signal in the Spectrogram plot.}
    \label{fig:d1}
\end{figure}

However, this problem is well suited for \texttt{LCDSC}. First, each infrasound is relatively quick (on the order of a few seconds within the 24-hour monitoring of the USArray sensors). Second, as seen in Figure \ref{fig:d1}, one of the canonical features of an infrasound dispersive wave is the presence of a ``gliding" or steadily increasing frequency in the signal. This makes infrasound dispersive waves display gliding similar to a Doppler signal reversed, which the \texttt{LCDSC} has performed well at cleaning. 
Performing \texttt{LCDSC} on the signal from one of the acoustic stations, we do indeed observe in Figure 9 that \texttt{LCDSC} cleans the signal well especially compared to WIT  which has made very little change to the signal due to the large period of noise throwing off the estimation WIT's baseline noise estimation.

\begin{figure}[!htbp]
    \centering
    \includegraphics[scale = 0.13]{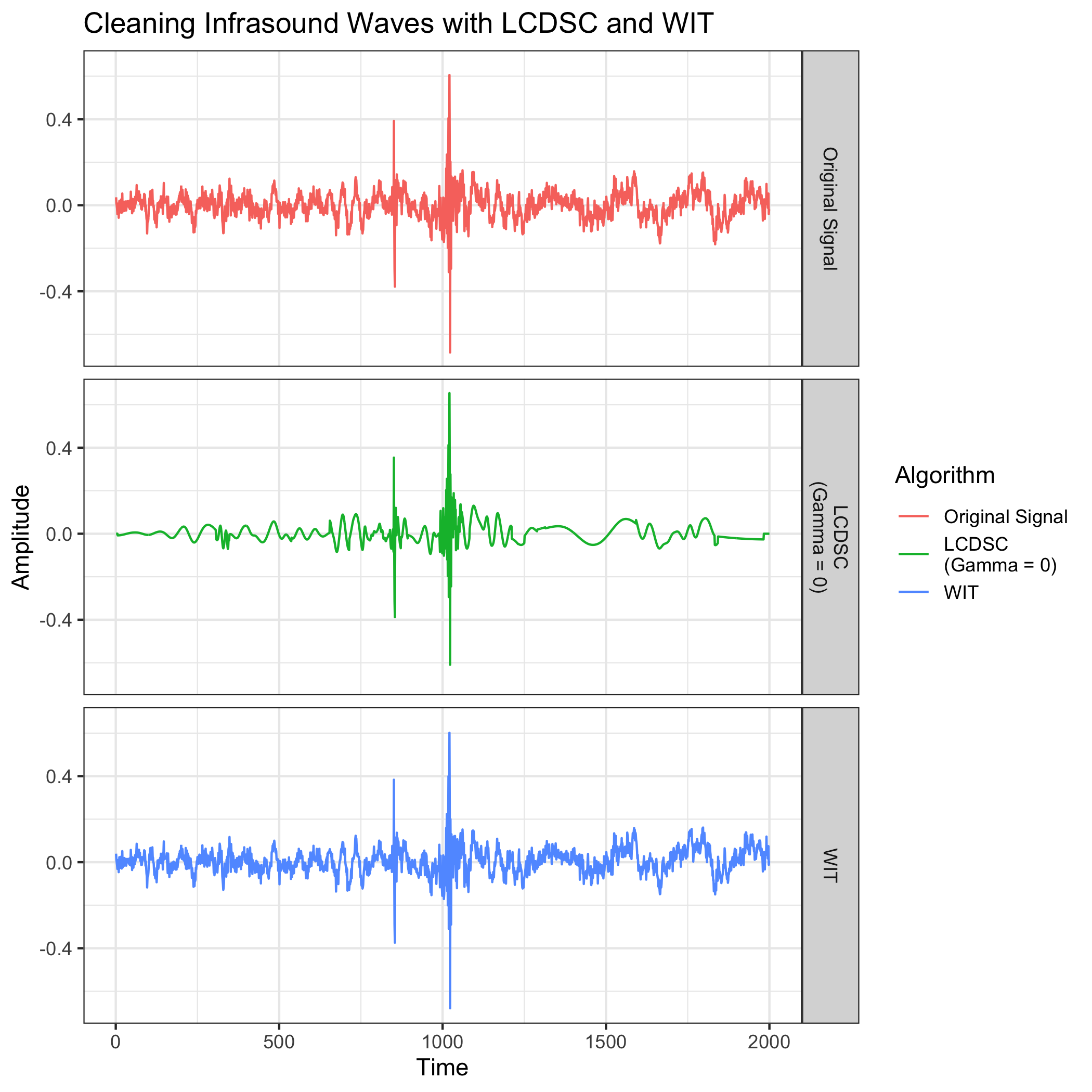}
    \label{fig:9}
    \caption{Comparison of uncleaned gliding signal with \texttt{LCDSC} cleaned signal and WIT cleaned signal. Note that \texttt{LCDSC} performs a better job at cleaning the signal than WIT, especially in helping to isolate the two spikes between 800-1000 that represent the infrasound dispersive wave.}
\end{figure}

Moreover, in Figure 10, by increasing the Threshold value, $\gamma$, we can clean the signal further and further until only the acoustic explosions are singled out. This occurs when gamma is around 2. This informs us that dispersive waves seem to lead to at least a 2 times increase in power in all of the IMFs.

\begin{figure}[!htbp]
    \centering
    \includegraphics[scale = 0.35]{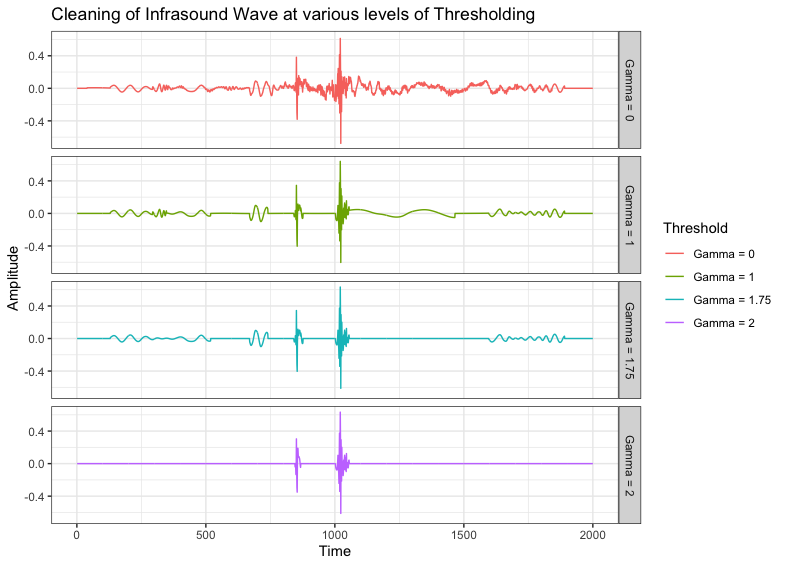}
    \label{fig:10}
    \caption{Comparison of \texttt{LCDSC} signal cleaning as $\gamma$ is increased. As $\gamma$ is increased, this results in a sparser and sparser signal cleaning, with $\gamma=2$ most cleanly isolating the dispersive wave. This indicates that the dispersive wave can be identified in each IMF as a 2 fold increase in SNR compared to background noise.}
\end{figure}

\section{Conclusion}
Here we provided a demonstration of the utility of \texttt{LCDSC} for the problem of local change point detection and signal cleaning. While other EEMD signal cleaning algorithms can exhibit drawbacks when there are long periods of no signal, our \texttt{LCDSC} does not suffer from the same disadvantage. This makes it ideal for the cleaning of short-term signals such as acoustic shock waves. We believe that the future development of EEMD signal decomposition will benefit greatly from the further development of methods based on local changes in basis functions.

\newpage
\section{Declarations}
Partial Financial Support for Authors Kentaro Hoffman and Kai Zhang has been provided by National Science Foundation under Grant No. DMS-1613112, IIS-1633212, DMS-1916237  , and DMS-1929298. Author Jonathan Lees did not receive support from any organization for the submitted work. All authors have no competing interests to declare that are relevant to the content of this article.

\section{Data Availability}
The datasets generated during and/or analyzed during the current study are available from the corresponding author on reasonable request.
\newpage


\bibliography{sn-bibliography}


\end{document}